\documentclass[aps,prd,reprint,twocolumn,tightenlines,superscriptaddress,floatfix,nofootinbib,showkeys]{revtex4-1}
\usepackage{XCharter}
\usepackage[T1]{fontenc}
\usepackage{fullpage}
\usepackage{amsfonts}
\usepackage{amsmath}
\usepackage{slashed}
\usepackage{amssymb}
\usepackage{comment}
\usepackage{graphicx}
\usepackage{epic}
\usepackage{eepic}
\usepackage{epsfig}
\usepackage{latexsym}
\usepackage{color}
\usepackage{float}
\usepackage{multirow}
\usepackage{hyperref}
\hypersetup{colorlinks=true,citecolor=cyan,urlcolor=blue,bookmarks=true,bookmarks=true,bookmarksopen=true,bookmarksnumbered=true,bookmarksopenlevel=3}
\usepackage{enumitem}
\usepackage[caption=false]{subfig}
\usepackage{natbib}
\usepackage{relsize}
\usepackage[left=2cm,right=2cm,top=1.9cm,bottom=1.95cm]{geometry}
\usepackage{diagbox}
\usepackage{ulem}
\usepackage{xcolor}
\usepackage{soul}

\renewcommand{\thefootnote}{\fnsymbol{footnote}}

\newcommand{\replyhe}[1]{{\color{black} #1}}

\begin{document}

\title{The role of triangle singularity in the $B^0 \to D^- \pi^+ a_0(980)(\pi^0 \eta)$ decay}

\author{Wei Wang}
\affiliation{School of Physics, Dalian University of Technology, \\ No.2 Linggong Road, Dalian, Liaoning, 116024, P.R.China}

\author{Dazhuang He}
\affiliation{College of Physics and Electronic Engineering, Heze University,\\ No.2269 University Road, Heze, Shandong, 274015, P.R.China}

\author{Xuan Luo}
\affiliation{School of Physics and Optoelectronics Engineering, Anhui University,\\ Hefei, Anhui 230601, People’s Republic of China}
\footnotetext[3]{xuanluo@ahu.edu.cn}

\begin{abstract}
The triangle singularity interpretation of the \replyhe{$a_1(1420)$} observed by the COMPASS Collaboration has been widely accepted.
In this work, we investigate the triangle mechanism in the decay $B^0 \to D^- \pi^+ a_0(980)(\pi^0 \eta)$, where the $a_0(980)$ is treated as a dynamically generated state.
The $\bar{K}^{*0}$-$K^+$-$K^-$ triangle loop originates from the decay $B^0 \to D^- K^+ \bar{K}^{*0}$, which has been observed by the Belle Collaboration, followed by the subsequent decay $\bar{K}^{*0} \to K^- \pi^+$.
The triangle amplitude develops a pronounced peak around 1420 MeV, which is reflected in the invariant mass spectrum of the $\pi a_0(980)$ system.
The differential decay width is calculated and exhibits a narrow peak around $980~\mathrm{MeV}$ in the $\pi^0 \eta$ invariant mass distribution.
Furthermore, the invariant mass distribution of the $\pi^+ \pi^0 \eta$ system shows a clear peak around $1420~\mathrm{MeV}$, which further \replyhe{confirms} the triangle singularity explanation of the \replyhe{$a_1(1420)$}.
We expect that the proposed $B^0$ decay mode could provide a potential platform for further exploring the triangle-singularity nature of the \replyhe{$a_1(1420)$} and could be tested in future experiments such as LHCb, BESIII, and Belle~II.
\end{abstract}

\renewcommand{\thefootnote}{\arabic{footnote}}
\setcounter{footnote}{0} 
\maketitle
\section{Introduction}
\label{sec:intro}
Over the past few decades, significant progress has been made in studies of the hadron spectrum, driven by the observation of numerous exotic states~\cite{ParticleDataGroup:2024cfk}.
Although many of these exotic states can be interpreted as multiquark states or hadronic molecules, some observed structures may not correspond to poles of the $S$-matrix but may instead arise from kinematic singularities~\cite{Guo:2017jvc,Olsen:2017bmm,Brambilla:2019esw}.
Among such kinematic effects, the triangle singularity (TS), first proposed by Landau in 1959~\cite{Karplus:1958zz,Landau:1959fi}, has attracted increasing interest. 
A triangle mechanism originates from the following sequential process:
an initial particle $A$ decays into two internal particles, labeled $1$ and $2$, which move back-to-back in the rest frame of $A$.
Particle $2$ subsequently decays into an internal particle $3$ and an external particle $B$, with particle $3$ moving in the same direction as particle $1$.
The two internal particles $1$ and $3$ then undergo rescattering and form an external particle $C$.
According to the Coleman-Norton theorem~\cite{Coleman:1965xm}, the emergence of a TS depends on whether the above processes can be interpreted as a classical scattering process and whether all three internal particles can simultaneously go on shell and become collinear in the rest frame of the decaying particle \cite{Karplus:1958zz}. 
In reality, the internal particles have finite widths, which smear the singular behavior and transform the TS into a finite peak that can be observed experimentally.

Phenomenological studies have successfully applied the TS mechanism to explain a variety of long-standing puzzles in hadron physics.
The anomalously large isospin violation observed in the decay $J/\psi \to \gamma \eta(1405/1475) \to \gamma \pi^0 f_0(980) \to \gamma 3\pi$, reported by the BESIII Collaboration~\cite{Li:2011ve}, was interpreted in terms of the TS in Ref.~\cite{Wu:2011yx}.
This mechanism has contributed to a better understanding of the nature of the two nearby states $\eta(1405)$ and $\eta(1475)$, as well as the mixing between the $a_0(980)$ and $f_0(980)$ resonances, which has been further investigated in a series of subsequent works~\cite{Aceti:2012dj,Wu:2012pg,Achasov:2015uua,Aceti:2015zva,Mikhasenko:2015oxp}.
Meanwhile, the triangle mechanism has also been employed to interpret the properties of heavy exotic hadronic states, such as the $Y(4260)$, $Z_c(3900)$, $Z_b(10610)$, and $Z_b(10650)$, in various processes~\cite{Guo:2011dv,Wu:2012ef,Li:2012as,Liu:2013uwx,Wang:2013cya,Wang:2013hga,Liu:2013vfa,Wu:2013onz,Liu:2014spa}.
A milestone in the development of the triangle mechanism was achieved in Ref.~\cite{Bayar:2016ftu}, where a detailed analysis of the singularities in a triangle loop integral was presented and a compact formula was derived for evaluating the TS on the physical boundary.
This formalism was successfully applied to the decay $\Lambda_b \to J/\psi K^- p$ through a $\Lambda^*$-charmonium-proton triangle loop.
Thereafter, the triangle mechanism involving an $f_0(980)$ or $a_0(980)$ final state was further explored in a variety of processes~\cite{Achasov:2016wll,Pavao:2017kcr,Aceti:2016yeb,Sakai:2017iqs,Liang:2017ijf,Achasov:2018swa,Lu:2022kdh}.
In Ref.~\cite{Pavao:2017kcr}, the authors identified two nonresonant peaks associated with the TS at approximately $2850$ and $3000~\mathrm{MeV}$ in the invariant mass distributions of the $\pi D_{s0}$ and $\pi D_{s1}$ systems, respectively.
That work also demonstrated the relation between the structure of the triangle amplitude and the finite widths of the internal particles in the loop.
A comprehensive review of threshold cusps and various TS structures in hadronic reactions can be found in Ref.~\cite{Guo:2019twa}, where their roles in phenomena related to exotic hadron candidates are systematically summarized.

The triangle mechanism has been extensively discussed in a wide range of hadronic decays involving charmonium~\cite{Jing:2019cbw,Liang:2019jtr,Huang:2021olv,Huang:2024oai,Xiao:2024ohf,Li:2025rlj,Wang:2023xua},
the $\Lambda$ baryon family~\cite{Sakai:2020fjh,Xie:2017mbe,Dai:2018hqb,Li:2025yad,Zhang:2024jby,Wang:2024ewe,Xie:2018gbi},
and the $D$-meson sector~\cite{Ling:2021qzl,Hsiao:2019ait,Bayar:2023azy,Ding:2020dio}.
In addition, decays of the $B$-meson family provide a promising platform for probing this mechanism, such as
$B^- \to D^{*0} \pi^- \pi^+ \pi^-(\pi^0 \eta)$~\cite{Pavao:2017kcr}, 
$B^- \to K^- \pi^0 X(3872)$~\cite{Sakai:2020ucu}, 
$B^- \to K^- X(3872)$ with $X(3872) \to \pi^0 \pi^+ \pi^-$~\cite{Molina:2020kyu}, 
$B^- \to K^- \pi^- D_{s0/s1}^+$~\cite{Sakai:2017hpg}, 
$B^+ \to J/\psi \phi K^+$~\cite{Ge:2021sdq}, 
$B^+ \to J/\psi \pi^{+} \pi^0 K^0$ and $B^+ \to J/\psi \pi^{+} \pi^- K^+$~\cite{Ren:2019rts}, 
$B^+ \to D^- D_s^+ \pi^+$ and $B^0 \to \bar{D}^0 D_s^+ \pi^-$~\cite{Ge:2022dsp}, 
$B^0 \to J/\psi K^0 f_0(980)(a_0(980))$~\cite{Lu:2022kdh}, 
$\bar{B}^0 \to \chi_{c1} K^- \pi^+$~\cite{Nakamura:2019emd}, 
$B \to (J/\psi \pi^+ \pi^-) K \pi$~\cite{Nakamura:2019nwd}, 
$\bar{B}_s^0 \to J/\psi \pi^0 f_0(980)$~\cite{Liang:2017ijf}, 
and $B_c \to B_s \pi \pi$~\cite{Samart:2017scf}.
Beyond hadronic decays, the triangle mechanism has also been explored in a variety of other processes, including semileptonic $\tau$-lepton decays~\cite{Dai:2018rra,Oset:2018zgc,Dai:2018zki},
photon-proton collisions~\cite{Wang:2016dtb,Debastiani:2017dlz}, electron-positron annihilation~\cite{Wang:2025zbv,Wei:2025ejv}, and proton-proton collisions~\cite{Bayar:2017svj,Ikeno:2021frl}.
Furthermore, an interesting study~\cite{Gao:2022jcy} discussed the structure of TS involving new-physics particles at high-energy colliders.
In addition, numerous other investigations have been devoted to the triangle mechanism and its phenomenological implications~\cite{Xie:2016lvs,Huang:2018wth,Roca:2017bvy,Liu:2019dqc,Xie:2019iwz,Debastiani:2018xoi,Huang:2020ptc,Huang:2018wgr,Liu:2015taa,Li:2025ejt,Achasov:2024anu,Shen:2020gpw}.

In particular, the TS explanation for the observation of the $a_1(1420)$ reported by the \replyhe{COMPASS Collaboration}~\cite{COMPASS:2015kdx,COMPASS:2020yhb} has been widely accepted.
In this mechanism, the $a_1(1260)$ first decays into $K^{*}\bar{K}$, with $K^{*}\to \pi K$, after which the $K\bar{K}$ pair fuses to form the $f_0(980)$, giving the observed $\pi f_0(980)$ decay mode.
\replyhe{In addition}, it has also been studied in various processes~\cite{Sakai:2017iqs,Liang:2017ijf,Dai:2018rra,Sakai:2020fjh,Molina:2021awn,Xiao:2024ohf}.
Based on previous studies, in this work we explore the physical effects of the TS and identify its contribution to the $B^0 \to D^- \pi^+ a_0(980)$ decay.
The observation of the weak decay $B^0 \to D^- K^+ \bar{K}^{*0}$ by the Belle Collaboration~\cite{Belle-II:2024xtf}, together with the cascade decay $\bar{K}^{*0} \to K^- \pi^+$, allows the construction of a $\bar{K}^{*0}$-$K$-$K$ triangle loop.
As we will show, the invariant mass of the $\bar{K}^{*0} K^+$ system in the $B^0$ decay allows the production of the $a_0(980)$ in the region where the TS condition is satisfied.
This leads to a pronounced peak structure in the $\pi^+ a_0(980)$ invariant mass spectrum around $1420~\mathrm{MeV}$.
In the above calculation, the triangle amplitude is evaluated following the analytical formulations developed in Refs.~\cite{Bayar:2016ftu,Guo:2019twa}.
Furthermore, within the chiral unitary approach, the $a_0(980)$ can be regarded as a dynamically generated state arising from meson-meson interactions~\cite{Liang:2014tia,Oller:1997ti,Oller:1998hw}.
An enhancement around $980~\mathrm{MeV}$ is expected to be reproduced in the $\pi^0 \eta$ invariant mass distribution.
Consequently, the $B^0 \to D^- \pi^+ a_0(980)(\pi^0\eta)$ decay provides a potential platform to investigate both the TS structure associated with the $\bar{K}^{*0}$-$K$-$K$ triangle loop and the nature of the $a_0(980)$ resonance.

The paper is organized as follows.
In Sec.~\ref{sec:Formalism}, we present the theoretical formalism and detail the kinematic conditions for the $\bar{K}^{*0}$-$K$-$K$ triangle loop considered here. 
In Sec.~\ref{sec:decay1}, the weak decay mechanism for $B^0 \to D^- K^+ \bar{K}^{*0}$ is described, and the effective coupling strength is extracted from current experimental measurements.
In Sec.~\ref{sec:decay2}, we formulate the $B^0 \to D^- \pi^+ a_0(980)$ decay by incorporating the triangle mechanism and provide the details of the derivation.
The $a_0(980)$ is further treated as a dynamically generated state, and the sequential process $B^0 \to D^- \pi^+ a_0(980)$ with $a_0(980) \to \pi^0 \eta$ is analyzed.
In Sec.~\ref{sec:res}, we present the numerical results for these processes.
The peaks around $1420$ and $980~\mathrm{MeV}$ are reproduced in the invariant mass distributions of the \replyhe{$\pi a_0$ and $\pi^0 \eta$ systems}, respectively.
Finally, a brief summary is provided in Sec.~\ref{sec:sum}.

\section{Formalism}
\label{sec:Formalism}
\begin{figure}[t]
\centering
\includegraphics[trim = 18 469 18 18, scale=0.22]{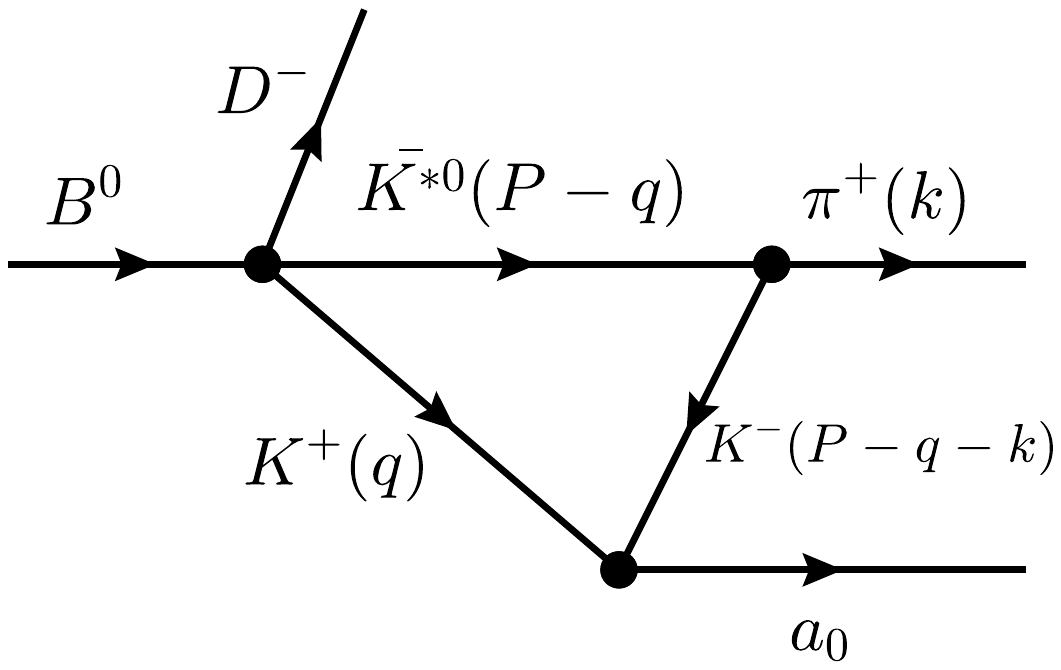}
\includegraphics[trim = 18 444 18 18, scale=0.2]{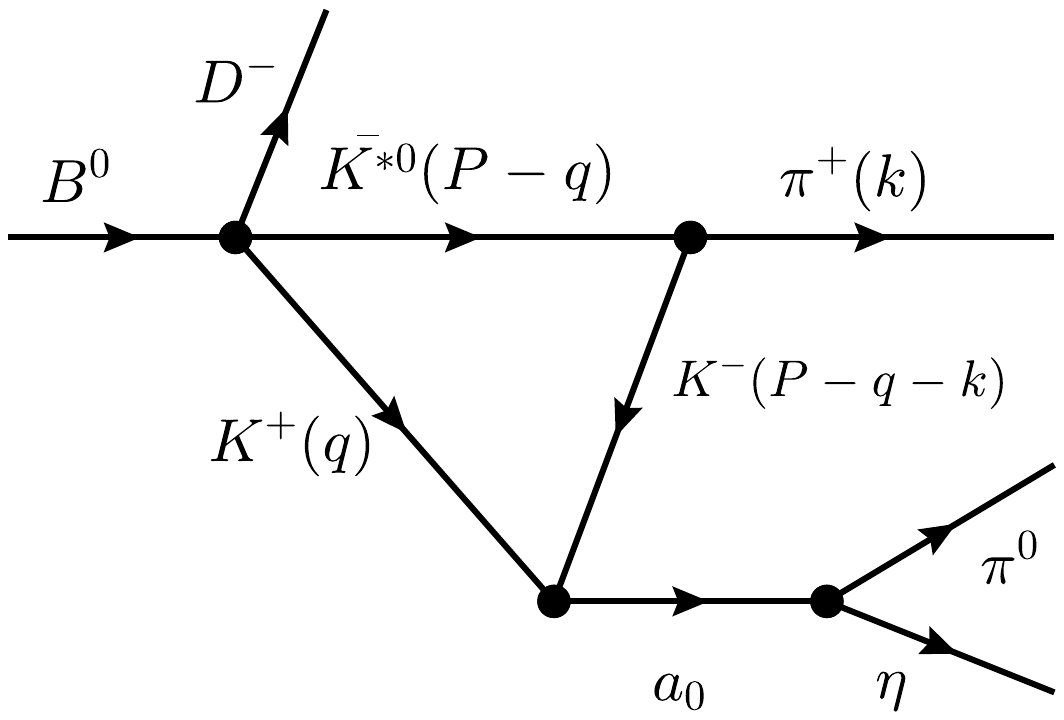}
\caption{The Feynman diagrams for the $B^0 \to D^- \pi^+ a_0(980)$ and $B^0 \to D^- \pi^+ a_0(980), a_0(980)\to\pi^0 \eta$ processes involving a $\bar{K}^{*}$-$K$-$\bar{K}$ triangle loop are shown in the left and right panels, respectively.}
\label{fig:B0a0}
\end{figure}

In this section, we show that a peak structure around $1420~\mathrm{MeV}$, induced by the TS, can be generated in the $B^0 \to D^- \pi^+ a_0(980)$ decay.
The Feynman diagram is shown in the left panel of Fig.~\ref{fig:B0a0}.
\replyhe{In this process, the $B^0$ first decays} into $D^- \bar{K}^{*0} K^+$, after which the $ \bar{K}^{*0}$ decays into $K^-$ and $\pi^+$.
The $K^+$ and $K^-$ move in the same direction, with the $K^-$ moving faster than the $K^+$, allowing the $K^+K^-$ pair to fuse into the $a_0(980)$.
The channel $B^0 \to D^{-} \bar{K}^{0} K^{*+}$ has not yet \replyhe{been observed experimentally}.
Therefore, the contribution of the $K^{*+}$-$\bar{K}^{0}$-$K^{0}$ triangle loop from the $B^0 \to D^{-} \bar{K}^{0} K^{*+}$ decay is not considered in the present work.
Furthermore, the $a_0(980)$ can be interpreted as a dynamically generated state with isospin $I=1$, arising from the coupled channels $\pi^0\eta$, $K^+K^-$, and $K^0\bar{K}^0$ within the chiral unitary approach~\cite{Oller:1997ti,Oller:1998hw}.
The decay of $a_0(980)$ is shown in the right panel of Fig.~\ref{fig:B0a0}.

The TS occurs when the kinematic \replyhe{variables} of the internal particles satisfy the following condition
\begin{equation}\label{Eq:2-1}
\begin{aligned}
\lim_{\epsilon \to 0}&\left( q_{+}^{\text{on}} -q_-^{\text{a}} \right)=0, \\
q_{+}^{\text{on}}&=\frac{\lambda^{\frac{1}{2}}\left(m_{\text{inv}}^2(\pi a_0),m_{K}^2,m_{\bar{K}^{*0}}^2\right)}{2m_{\text{inv}}(\pi a_0)} + i \epsilon,
\end{aligned}
\end{equation}
where $q_{+}^{\text{on}}$ denotes the on-shell three-momentum of $K^{+}$ in the center-of-mass frame (COM) of the $\bar{K}^{*0} K^+$ system.
The quantity $m_{\text{inv}}(\pi a_0)$ is the invariant mass of the $\bar{K}^{*0} K^+$ system, and $\lambda(x, y, z)=x^2 +y^2 +z^2 -2xy-2yz -2xz$ is the K\"{a}ll\'{e}n function. 
Meanwhile, $q_-^{\text{a}}$ is given by
\begin{equation}\label{Eq:2-2}
\begin{aligned}
q_-^{\text{a}} =\gamma(\nu E_{K^{-}}^*-p_{K^{-}}^*)-i \epsilon,
\end{aligned}
\end{equation}
with the definitions
\begin{equation}\label{Eq:2-3}
\begin{aligned}
\nu =& \frac{k}{E_{a_0}}, \qquad   \gamma=\frac{1}{\sqrt{1-\nu^2}}=\frac{E_{a_0}}{m_{a_0}},\\
E_{K^{-}}^*=&\frac{m_{a_0}}{2},\qquad p_{K^{-}}^*=\frac{\lambda^{\frac{1}{2}}(m_{a_0}^2,m_{K}^2,m_{K}^2)}{2m_{a_0}},
\end{aligned}
\end{equation}
where $E_{K^{-}}^*$ and $p_{K^{-}}^*$ are the energy and momentum of the $K^{-}$ meson in the rest frame of $a_0$.
$\nu$ and $\gamma$ denote the velocity of the $K^+ K^-$ system and the Lorentz boost factor, respectively.
Eq.~\ref{Eq:2-1} implies that all three particles in the loop are on shell and that $K^+$ in the rest frame of $a_0(980)$ and $a_0(980)$ in the COM frame of $\pi a_0(980)$ move in the same direction.
The formation of $a_0(980)$ through $K^+ K^-$ fusion further requires that the momentum of $K^+$ in the COM frame of $\pi  a_0$ be smaller than that of $K^-$ in the rest frame of $\bar{K}^{*0}$.
Equivalently, Eq.~\ref{Eq:2-1} can be understood as the condition that $q^a_-$ and $q^{\text{on}}_+$ represent the singularities of the triangle loop function in the upper and lower halves of the complex-$q$ plane, respectively.
The integration contour of the loop function in Eq.~\ref{Eq:b9} is then pinched between $q^a_-$ and $q^{\text{on}}_+$ at the same point on the real axis.
Consequently, solving Eq.~\ref{Eq:2-1} yields a TS around $1420$ MeV in the $m_{\text{inv}}(\pi a_0)$ invariant mass distribution.
When the finite width of the $\bar{K}^{*0}$ is taken into account by replacing $m_{\bar{K}^{*0}} \to m_{\bar{K}^{*0}} - i\Gamma_{\bar{K}^{*0}}/2$ with $\Gamma_{\bar{K}^{*0}} = 47~\mathrm{MeV}$, the TS moves into the complex plane, yielding $1420.7 - i\,28.0~\mathrm{MeV}$.

\section{the $B^0 \to D^- K^+ \bar{K}^{*0}$ decay}
\label{sec:decay1}

\begin{figure}[!ht]
\centering
\includegraphics[trim = 18 596 18 18, scale=0.4]{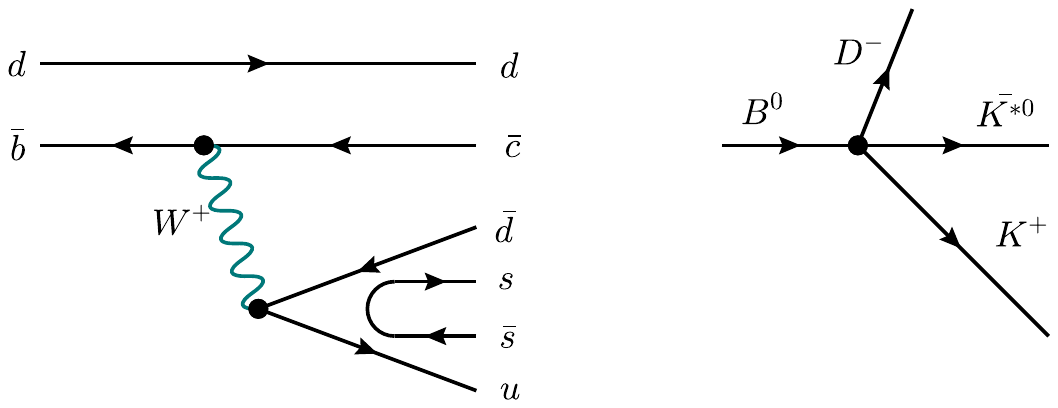}
\caption{The quark-level and hadron-level Feynman diagrams for the decay $B^0 \to D^- K^+ \bar{K}^{*0}$ are shown in the left and right panels, respectively.}
\label{fig:Bdecay1}
\end{figure}
To evaluate the amplitude shown in Fig.~\ref{fig:B0a0}, the effective coupling strength of the $B^0 \to  D^- K^+ \bar{K}^{*0}$ vertex is required.
The $B^0$ meson can decay into the $D^- K^+ \bar{K}^{*0}$ final state through the weak decay of the $b$ quark, as illustrated in Fig.~\ref{fig:Bdecay1}.
At the quark level, the $b$ quark undergoes a weak transition into a $c$ quark through the emission of a $W$ boson and hadronizes into the $D^-$ meson.
The emitted $W$ boson subsequently decays into a $u\bar{d}$ pair, which combines with an additional $q\bar{q}$ pair from the vacuum to form the $\bar{K}^{*0}K^+$ final state.

The interaction of the effective vertex in Fig.~\ref{fig:Bdecay1} can be constructed using the $P$-wave interaction.
Following the convention adopted in Refs.~\cite{Sakai:2017iqs,Sakai:2017hpg}, we write  
\begin{equation}\label{Eq:a1}
\begin{aligned}
-it_{B \to D K \bar{K}^{*0}}=-iC \vec{\epsilon}_{\bar{K}^{*0}} \cdot \vec{p}_{D},
\end{aligned}
\end{equation}
where $\vec{\epsilon}_{\bar{K}^{*0}}$ and $ \vec{p}_{D}$ are the polarization vector of the $\bar{K}^{*0}$ and the momentum of the $D^-$, respectively. 
Below, we present the details of the calculation of the vertex coupling strength $C$.
By solving Eq.~\ref{Eq:2-1}, a TS around $1420~\mathrm{MeV}$ is obtained.
The three-momentum of the $\bar{K}^{*0}$ in the $\pi a_0$ rest frame is approximately $135.66~\mathrm{MeV}$, which is smaller than the $\bar{K}^{*0}$ mass of $895.81$ MeV. 
Therefore, the time component $\epsilon^0$ of the $\bar{K}^{*0}$ polarization vector can be safely neglected.
We take the polarization sum for the $\bar{K}^{*0}$ as
\begin{equation}\label{Eq:a2}
\begin{aligned}
\sum_{\mu, \nu} &\epsilon_{\bar{K}^{*0}\mu}\epsilon_{\bar{K}^{*0}\nu} \sim \sum_{i,j} \epsilon_{\bar{K}^{*0}i}\epsilon_{\bar{K}^{*0}j}=\delta_{ij}; \\
&\mu=i,\ \mu=j;\qquad i,j=1,2,3.
\end{aligned}
\end{equation}
The decay width for the process $B \to D K \bar{K}^{*0}$ is given by
\begin{equation}\label{Eq:a3}
\begin{aligned}
\Gamma_{B \to D K \bar{K}^{*0}} &=\int dm_{\text{inv}}( \bar{K}^{*0} K^+) \\
&\frac{1}{(2\pi)^3}\frac{|\vec{\tilde{p}}_{D} | |\vec{\tilde{p}}^\prime_{\bar{K}^{*0}} | }{4m_{B}^2} \sum_{\text{pol}}|t_{B \to D K \bar{K}^{*0}}|^2, \\
\end{aligned}
\end{equation}
where $m_{\mathrm{inv}}(\bar{K}^{*0}K^+)$ is the invariant mass of the $\bar{K}^{*0}K^+$ system.
Here, $\vec{\tilde{p}}^{\,\prime}_{\bar{K}^{*0}}$ and $\vec{\tilde{p}}_{D}$ denote the three-momenta of the $\bar{K}^{*0}$ in the $\bar{K}^{*0}K^+$ COM frame and of the $D^-$ in the $B^0$ rest frame, respectively.
They are given by
\begin{equation}\label{Eq:a4}
\begin{aligned}
|\vec{\tilde{p}}_{D} |&=\frac{\lambda^{\frac{1}{2}}\left(m^2_{B},\ m_{\text{inv}}^2( \bar{K}^{*0} K^+),\ m^2_{D} \right)}{2 m_{B}}, \\
|\vec{\tilde{p}}^\prime_{\bar{K}^{*0}} | &=\frac{\lambda^{\frac{1}{2}}\left(m_{\text{inv}}^2( \bar{K}^{*0} K^+),\ m^2_{K},\ m^2_{\bar{K}^{*0}}\right)}{2 m_{\text{inv}}( \bar{K}^{*0} K^+ )}.\\
\end{aligned}
\end{equation}
After squaring the amplitude $t_{B \to D K \bar{K}^{*0} }$ and applying the polarization sum in Eq.~\ref{Eq:a2}, we obtain
\begin{equation}\label{Eq:a5}
\sum _{\text{pol}} |t_{B \to D K \bar{K}^{*0}}|^2=C^2 |\vec{\tilde{p}}_{D}^\prime|^2,
\end{equation}
where $\vec{\tilde{p}}_{D}^\prime$ is the three-momentum of the $D^-$ in the COM frame of the $\bar{K}^{*0}K^+$ system.
\begin{equation}
|\vec{\tilde{p}}_{D}^\prime |=\frac{\lambda^{\frac{1}{2}}\left(m^2_{B},\ m_{\text{inv}}^2( \bar{K}^{*0} K^+),\ m^2_{D} \right)}{2 m_{\text{inv}}( \bar{K}^{*0} K^+)}.
\end{equation}

From recent measurements~\cite{ParticleDataGroup:2024cfk,Belle-II:2024xtf}, the partial decay branching fraction is given by Br($B^0 \to D^- K^+ \bar{K}^{*0}$) = $(7.7\pm 0.6)\times 10^{-4}$.
Combining Eq.~\ref{Eq:a3} with Eq.~\ref{Eq:a5}, we obtain
\begin{equation}\label{Eq:a6}
\begin{aligned}
\frac{C^2}{\Gamma_{B^0}}=\frac{\text{Br} \left( B^0 \to D^- K^+ \bar{K}^{*0}\right) }{\int dm_{\text{inv}}( \bar{K}^{*0} K^+)\frac{1}{(2\pi)^3}\frac{ |\vec{\tilde{p}}_{D} | |\vec{\tilde{p}}^\prime_{\bar{K}^{*0}} |}{4m_{B}^2} |\vec{\tilde{p}}_{D}^\prime|^2}.
\end{aligned}
\end{equation}

\section{the $B^0 \to D^- \pi^+ a_0 (980)$ decay}
\label{sec:decay2}

The amplitude for the $B^0 \to D^- \pi^+ a_0(980)$ decay (left panel of FIG.~\ref{fig:B0a0}) can be written as 
\begin{equation}\label{Eq:b1}
\begin{aligned}
&-it_{B \to D \pi a_0}=i\sum_{\text{pol}} \int \frac{d^4 q}{(2\pi)^4}\ \frac{it_{B \to D K \bar{K}^{*0}}}{q^2-m_{K}^2+i \epsilon}\\
&\frac{it_{\bar{K}^{*0} K^- \pi^+}}{(P-q)^2-m_{\bar{K}^{*0}}^2+i \epsilon}\  \frac{it_{K^+ K^- a_0 } }{(P-q-k)^2-m_{K}^2+i \epsilon},
\end{aligned}
\end{equation}
in the COM frame of the $\pi a_0$ system\replyhe{, where} the $\pi^+$ originates from the $\bar{K}^{*0}$ decay.
In Eq.~\ref{Eq:b1}, the amplitude $t_{B \to D K \bar{K}^{*0}}$ was calculated in the previous section (Sec.~\ref{sec:decay1}).
The amplitude $t_{\bar{K}^{*0} K^- \pi^+}$ can be calculated using the chiral-invariant Lagrangian with local hidden symmetry~\cite{Scherer:2002tk,Pich:1995bw} and is given by
\begin{equation}\label{Eq:b2}
\mathcal{L}_{\text{VPP}}=-ig\langle V^\mu[P,\partial_\mu P]\rangle,
\end{equation}
where \replyhe{bracket} $\langle \cdots \rangle$ denotes the SU(3) trace, and the coupling constant is given by $g=m_V/2f_\pi$, with $m_V=800$ MeV and $f_\pi=93$ MeV, within the local hidden gauge formalism. 
$P$ and $V^\mu$ represent the pseudoscalar and vector meson octets, respectively, which are given by
\begin{equation}\label{Eq:b3}
\begin{aligned}
&P=\begin{pmatrix}
\frac{\pi^0}{\sqrt{2}}+\frac{\eta_8}{\sqrt{6}} &\pi^+ &K^+\\
 \pi^-&-\frac{\pi^0}{\sqrt{2}}+\frac{\eta_8}{\sqrt{6}} &K^0\\
 K^-& \bar{K}^0& -\frac{2}{\sqrt{6}}\eta_8
\end{pmatrix}_, \\
&V=\begin{pmatrix}
	\frac{\rho^0}{\sqrt{2}}+\frac{\omega}{\sqrt{2}}& \rho^+& K^{*+}\\
	\rho^-&-\frac{\rho^0}{\sqrt{2}}+\frac{\omega}{\sqrt{2}} & K^{*0}\\
	K^{*-}& \bar{K}^{*0}& \phi
	\end{pmatrix}_.
\end{aligned}
\end{equation}
Then, the amplitude for the decay $\bar{K}^{*0} \to K^- \pi^+$ can be written as
\begin{equation}\label{Eq:b4}
-it_{\bar{K}^{*0} K^- \pi^+}=-ig\vec{\epsilon}_{\bar{K}^{*0}} \cdot (\vec{p}_{\pi^+}^{\,\prime}-\vec{p}_{K^-}^{\,\prime}).
\end{equation}
As in Eq.~\ref{Eq:a5}, we neglect the time component of the polarization vector in Eq.~\ref{Eq:b4} and employ the polarization sum given in Eq.~\ref{Eq:a2}.
Here, $\vec{p}_{\pi^+}^{\,\prime}$ and $\vec{p}_{K^-}^{\,\prime}$ denote the three-momenta of the $\pi^+$ and $K^-$ in the COM frame of the $\pi a_0$ system, respectively.
The momentum $\vec{p}_{\pi^+}^{\,\prime}$ is given by
\begin{equation}\label{Eq:b5}
\begin{aligned}
|\vec{p}_{\pi^+}^{\,\prime}|&=|\vec{k}|=\frac{\lambda^{\frac{1}{2}}\left(m^2_{\text{inv}}(\pi a_0),\ m_{\pi}^2,\ m^2_{a_0} \right)}{2 m_{\text{inv}}(\pi a_0)}.\\
\end{aligned}
\end{equation}

As discussed in Sec.~\ref{sec:Formalism}, the $a_0(980)$ is considered to be a dynamically generated state.
The amplitude $t_{K^+ K^- a_0 }$ can be written directly as
\begin{equation}\label{Eq:b6}
t_{K^+ K^- a_0 }=g_{K^+ K^- a_0 }.
\end{equation}
Above, we set $g_{K^+ K^- a_0 } = 3875$ MeV~\cite{Oller:1997ti}.
Thus, the amplitude expression in Eq.~\ref{Eq:b1} can be \replyhe{simplified} to 
\begin{equation}\label{Eq:b7}
\begin{aligned}
&t_{B \to D \pi a_0 (980)}=i g_{K^+ K^- a_0 } g C \sum_{\text{pol}} \int \frac{d^4 q}{(2\pi)^4} \\
&\times \frac{\vec{\epsilon}_{\bar{K}^{*0}} \cdot \vec{p}_{D}^{\,\prime} }{q^2-m_{K}^2+i \epsilon}   \frac{\vec{\epsilon}_{\bar{K}^{*0}} \cdot (\vec{p}_{\pi^+}^{\,\prime}-\vec{p}_{K^-}^{\,\prime}) }{(P-q)^2-m_{\bar{K}^{*0}}^2+i \epsilon}  \\
&\times \frac{1}{(P-q-k)^2-m_{K}^2+i \epsilon} ,
\end{aligned}
\end{equation}
where $\vec{p}_{\pi^+}^{\,\prime}$ is given in Eq.~\ref{Eq:b5}. 
The quantity $\vec{p}_{D}^{\,\prime}$ is the momentum of the $D^-$ in the COM frame of $\pi a_0$, which originates from the $B^0$ decay.
\begin{equation}
|\vec{p}_{D}^{\,\prime}|=\frac{\lambda^{\frac{1}{2}}\left(m^2_{B},\ m_{\text{inv}}^2( \pi a_0),\ m^2_{D} \right)}{2 m_{\text{inv}}( \pi a_0)}.
\end{equation}
Equation~\ref{Eq:a2} is used to perform the polarization sum in Eq.~\ref{Eq:b7}, yielding
\begin{equation}\label{Eq:b8}
\begin{aligned}
t_{B \to D \pi a_0 (980)}&=i g_{K^+ K^- a_0 } g C  \int \frac{d^4 q}{(2\pi)^4} \\
\times &\frac{1 }{q^2-m_{K}^2+i \epsilon}   \frac{\vec{p}_{D}^{\,\prime} \cdot (2\vec{k}+\vec{q}) }{(P-q)^2-m_{\bar{K}^{*0}}^2+i \epsilon} \\
\times &\frac{1}{(P-q-k)^2-m_{K}^2+i \epsilon}, \\
\end{aligned}
\end{equation}
where $\vec{p}_{\pi^+}^{\,\prime}-\vec{p}_{K^-}^{\,\prime}=\vec{k}-(-\vec{k}-\vec{q})=2\vec{k}+\vec{q}$. 
$P$ is the momentum of the $\pi a_0$ system \replyhe{in its rest frame}. 
\begin{equation}\label{Eq:PP}
\begin{aligned}
P=\left(m_{\text{inv}}(\pi a_0),\ 0,\ 0,\ 0 \right).
\end{aligned}
\end{equation}
The quantity $t_T$ is defined to describe the loop integral in Eq.~\ref{Eq:b8} as
\begin{equation}\label{Eq:b9}
\begin{aligned}
&t_T=i\int \frac{d^4 q}{(2\pi)^4} \ \vec{p}_{D}^{\,\prime} \cdot (2\vec{k}+\vec{q})\frac{1 }{q^2-m_{K}^2+i \epsilon}  \\
&\frac{1 }{(P-q)^2-m_{\bar{K}^{*0}}^2+i \epsilon}  \frac{1}{(P-q-k)^2-m_{K}^2+i \epsilon}. \\
\end{aligned}
\end{equation}
Following Refs.\cite{Bayar:2016ftu,Guo:2019twa}, the negative-energy part of the $\bar{K}^{*0}$ propagator in $t_T$ can be neglected.
The integration over $dq^0$ in $t_T$ is performed by applying the residue theorem \replyhe{and using} the following formula
\begin{equation}\label{Eq:b10}
\int d^3q \ \vec{q}_i\  f(\vec{q},\vec{k})=\vec{k}_i \int d^3 q \frac{\vec{q}\cdot \vec{k}}{|\vec{k}|^2} f(\vec{q},\vec{k}).
\end{equation}
Now, the triangle loop amplitude $t_T$ reduces to
\begin{equation}\label{Eq:b11}
\begin{aligned}
t_T&=  \vec{p}_{D}^{\,\prime}  \cdot \vec{k}  \int \frac{d^3 q}{(2\pi)^3} {\left(2+  \frac{\vec{q}\cdot \vec{k}}{|\vec{k}|^2}\right) }\\
&\times \frac{1}{8\omega_{K^+}(\vec{q})\ \omega_{K^-}(\vec{q}+\vec{k})\ \omega_{K^*}(\vec{q})}\\
&\times \frac{1}{k^0 - \omega_{K^-}(\vec{q}+\vec{k}) - \omega_{K^*}(\vec{q}) + i\epsilon}\\
&\times \frac{1}{P^0 - \omega_{K^*}(\vec{q}) - \omega_{K^+}(\vec{q}) + i\epsilon}\\
&\times \frac{1}{P^0 - \omega_{K^+}(\vec{q}) - \omega_{K^-}(\vec{q}+\vec{k}) - k^0+i\epsilon}			\\
&\times \frac{1}{P^0 + \omega_{K^+}(\vec{q}) + \omega_{K^-}(\vec{q}+\vec{k}) - k^0 + i\epsilon}			\\
&\times \bigg\{ 2P^0\omega_{K^+}(\vec{q}) + 2k^0 \omega_{K^-}(\vec{q}+\vec{k}) \\
&- 2\left[ \omega_{K^+}(\vec{q}) + \omega_{K^-}(\vec{q}+\vec{k})\right] \\
&\times \left[ \omega_{K^+}(\vec{q}) + \omega_{K^-}(\vec{q}+\vec{k}) + \omega_{K^*}(\vec{q})\right] \bigg\},\\
&= \vec{p}_{D}^{\,\prime}  \cdot \vec{k} \times  \tilde{t}_T,
\end{aligned}
\end{equation}
where $\omega_{K^+}(\vec{q})=\sqrt{m_{K}^2 + \vec{q}^2}$, $\omega_{K^{*0}}(\vec{q})= \sqrt{m_{\bar{K}^{*0}}^2 + \vec{q}^2}$, and $\omega_{K^-}(\vec{q}+\vec{k}) = \sqrt{m_{K}^2 + (\vec{k}+\vec{q})^2}$.
From Eq.~\ref{Eq:PP}, $P^0 = m_{\mathrm{inv}}(\pi a_0)$ is the invariant mass of the $\pi a_0$ system.
Although the integral in Eq.~\ref{Eq:b11} is convergent, \replyhe{the chiral unitary approach naturally provides an upper limit $q_{\mathrm{max}}$}~\cite{Aceti:2015zva,Aceti:2012dj}.
In this work, we use $q_{\mathrm{max}} = 600~\mathrm{MeV}$ for $a_0(980)$ production~\cite{Liang:2014tia,Xie:2014tma}.
To reproduce the peak observed in experiments, the finite width of the $\bar{K}^{*0}$ in the triangle loop must be taken into account.
This is implemented by replacing $\omega_{K^{*}}$ with $\omega_{K^{*}} - i \Gamma_{\bar{K}^{*0}}/2$ in the denominator of Eq.~\ref{Eq:b11}.

Finally, the differential decay width for the three-body final state $B^0 \to D^- \pi^+ a_0(980)$ is given by
\begin{equation}\label{Gamma-1}
\begin{aligned}
\frac{d \Gamma^\prime}{d m_{\text{inv}}(\pi a_0)}&={\frac{2}{3}}\frac{1}{(2\pi)^3}\frac{ C^2g_{K^+ K^- a_0 }^2 g^2}{{8}m_{B}^2}	\\
	&\times {|\vec{p}_{D}| |\vec{p}_{D}^{\,\prime}|^2 |\vec{k}|^3 |\tilde{t}_T|^2},\\
\end{aligned}
\end{equation}
The factor $2/3$ in Eq.~\ref{Gamma-1} originates from the angular phase-space integration of the integrand in $\vec{p}_{D}^{\,\prime}  \cdot \vec{k}=|\vec{p}_{D}^{\,\prime}| | \vec{k}|\cos \theta$.
Here, $\vec{p}_{D}$ denotes the momentum of the $D^-$ in the $B^0$ rest frame.
\begin{equation}\label{Eq:b12}
|\vec{p}_{D}|=\frac{\lambda^{\frac{1}{2}}\left(m_{B}^2,M_{D^-}^2,m_{\text{inv}}^2(\pi a_0) \right)}{2m_{B}}.
\end{equation}

In addition, we further explore the possibility of observing the $a_0(980)$ through the $\pi^0 \eta$ final state.
Following Ref.~\cite{Sakai:2017hpg}, the $a_0(980)$ appears as an unstable resonance in the $B^0$ decay and subsequently decays into $\pi^0 \eta$.
Therefore, we consider the corresponding four-body decay, as illustrated in the right panel of Fig.~\ref{fig:B0a0}.
In this process, the $a_0(980)$ resonance can be treated as an intermediate particle, allowing us to include an additional differential distribution with respect to the $\pi\eta$ invariant mass in Eq.~\ref{Gamma-1}.
The differential decay width for $B^0 \to D^- \pi^+ \pi^0 \eta$ is then given by
\begin{equation}\label{Eq:b13}
\begin{aligned}
\frac{d \Gamma^\prime}{d m_{\text{inv}}(\pi a_0)}&=\frac{1}{3}\frac{1}{(2\pi)^3} \frac{g^2 C^2}{4m_{B}^2} \int dm_{\text{inv}}^2(\pi \eta) \\
&\left(-\frac{1}{\pi}\text{Im}D \right) g_{K^- K^+ a_0}^2 {|\vec{p}_{D}| |\vec{p}_{D}^{\,\prime}|^2 |\vec{k}|^3 |\tilde{t}_T|^2},\\
\end{aligned}
\end{equation}
with
\begin{equation}\label{Eq:b14}
D=\frac{1}{m_{\text{inv}}^2(\pi \eta)-m_{a_0}^2+im_{a_0}\Gamma_{a_0}},
\end{equation}
where $m_{\text{inv}}(\pi \eta)$ is the invariant mass of the $\pi \eta$ system produced in the $a_0(980)$ decay, and $m_{a_0}$ is the mass of $a_0(980)$. 
Eq.~\ref{Eq:b13} reduces to Eq.~\ref{Gamma-1} in the limit $\Gamma_{a_0} \to 0$, where $i\,\mathrm{Im}\,D \to -i\pi \, \delta\big(m_{\mathrm{inv}}^2(\pi \eta) - m_{a_0}^2 \big)$.
Thus, Eq.~\ref{Eq:b13} simplifies to
\begin{equation}\label{Eq:b15}
\begin{aligned}
\frac{d \Gamma^\prime}{d m_{\text{inv}}(\pi a_0)}&=\frac{1}{3 \pi}\frac{1}{(2\pi)^3} \frac{g^2 C^2}{4m_{B}^2} \int dm_{\text{inv}}^2(\pi \eta)\\
&\frac{g_{K^- K^+ a_0}^2 {|\vec{p}_{D}| |\vec{p}_{D}^{\,\prime}|^2 |\vec{k}|^3 |\tilde{t}_T|^2} m_{a_0} \Gamma_{a_0}}{\left[m_{\text{inv}}^2(\pi \eta)-m_{a_0}^2 \right]^2+(m_{a_0} \Gamma_{a_0})^2}.\\
\end{aligned}
\end{equation}
The decay width $\Gamma_{a_0}$ is calculated by considering the $a_0 \to \pi^0 \eta$ decay.
\begin{equation}\label{Eq:b16}
\begin{aligned}
\Gamma_{a_0}&=\frac{1}{8\pi}\frac{g_{a_0 \pi \eta}^2}{m_{\text{inv}}^2(\pi \eta)}|\vec{\tilde{q}}_{\eta}|, \\
\end{aligned}
\end{equation}
where $|\vec{\tilde{q}}_{\eta}|$ denotes the magnitude of the $\eta$ momentum in the rest frame of $a_0(980)$ and is given by
\begin{equation}\label{Eq:b17}
|\vec{\tilde{q}}_{\eta}|= \frac{\lambda^{\frac{1}{2}}\left( m_{\text{inv}}^2(\pi \eta),m_\pi^2,m_\eta^2\right)}{2m_{\text{inv}}(\pi \eta)}.
\end{equation}
By substituting Eq.~\ref{Eq:b16} into Eq.~\ref{Eq:b15}, \replyhe{we obtain}
\begin{equation}\label{Eq:b18}
\begin{aligned}
\frac{d \Gamma^\prime}{d m_{\text{inv}}(\pi a_0)}&=\frac{1}{(2\pi)^5} \frac{g^2 C^2}{24 m_{B}^2} \int dm_{\text{inv}}^2(\pi \eta)\\
&\times {|\vec{p}_{D}| |\vec{p}_{D}^{\,\prime}|^2 |\vec{k}|^3 |\tilde{t}_T|^2}  \frac{|\vec{\tilde{q}}_{\eta}|}{m_{\text{inv}}^2(\pi \eta)} \\
&\times \frac{m_{a_0} g_{K^- K^+ a_0}^2 g_{a_0 \pi \eta}^2}{\left[m_{\text{inv}}^2(\pi \eta)-m_{a_0}^2 \right]^2+(m_{a_0} \Gamma_{a_0})^2}.\\
\end{aligned}
\end{equation}
Meanwhile, \replyhe{the amplitude for the $K^+ K^- \to \pi \eta$ process mediated by exchange of the $a_0(980)$ resonance can be written as}
\begin{equation}\label{Eq:b19}
\frac{g_{K^- K^+ a_0}^2 g_{a_0 \pi \eta}^2}{\left[m_{\text{inv}}^2(\pi \eta)-m_{a_0}^2 \right]^2+(m_{a_0} \Gamma_{a_0})^2}=|t_{K^+ K^-\to \pi \eta}|^2.
\end{equation}
In contrast, when the $a_0(980)$ is treated as a dynamically generated state, the amplitude $t_{K^+ K^-\to \pi \eta}$ can be calculated by solving the Bethe--Salpeter (BS) equation within the chiral unitary approach. The BS equation is given by
\begin{equation}\label{Eq:b20}
T=\left[1-VG \right]^{-1}V.
\end{equation}
Here, $G$ denotes the loop function for two meson propagators~\cite{Liang:2014tia}.
\begin{equation}
\begin{aligned}
G(s) = i\int \frac{d^4 q}{(2\pi)^4} \frac{1}{(P-q)^2-m_1^2+i\epsilon} \frac{1}{q^2-m_2^2+i\epsilon}
\end{aligned}
\end{equation}
and is regularized by a cutoff $q_{\text{max}}$ as
\begin{equation}\nonumber
\begin{aligned}
&G(s) =  \frac{1}{16\pi^2 s} \Bigg\lbrace \sigma \left( \arctan \frac{s + \Delta}{\sigma \lambda_1} + \arctan \frac{s - \Delta}{\sigma \lambda_2} \right) \\
& - \left[ (s + \Delta) \ln \frac{(1 + \lambda_1) q_{\text{max}}}{m_1} + (s - \Delta) \ln \frac{(1 + \lambda_2) q_{\text{max}}}{m_2} \right] \Bigg\rbrace
\end{aligned}
\end{equation}
with
\begin{equation}\nonumber
\begin{aligned}
\sigma&=\left[-(s-(m_1+m_2)^2)(s-(m_1-m_2)^2)\right]^{1/2},\\
\Delta&=m_1^2-m_2^2,\lambda_1=\sqrt{1+\frac{m_1^2}{q^2_{\text{max}}}}, \lambda_2=\sqrt{1+\frac{m_2^2}{q^2_{\text{max}}}}
\end{aligned}
\end{equation}
A value of $q_{\text{max}}=600$ MeV is used to reproduce the $a_0(980)$ around 980 MeV.
The meson-meson interaction potentials for the case $I=1$ are given as follows~\cite{Liang:2014tia}:
\begin{equation}
\begin{aligned}
V_{K^{+}K^{-}\to K^{+}K^{-}} &= -\frac{1}{2f_\pi^2}s, \quad  V_{K^{+}K^{-}\to K^{0}\bar{K}^{0}} = -\frac{1}{4f_\pi^2}s,\\
V_{K^{+}K^{-}\to\pi^{0}\eta} &= \frac{- \sqrt{3}}{12f_\pi^2} \left( 3s - \frac{8}{3}m_K^2 - \frac{1}{3}m_{\pi}^2 - m_{\eta}^2 \right), \\
V_{K^{0}\bar{K}^{0}\to K^{0}\bar{K}^{0}} &= -\frac{1}{2f_\pi^2}s, \qquad \quad V_{\pi^{0}\eta\to\pi^{0}\eta} = -\frac{m_{\pi}^2}{3f_\pi^2}, \\
V_{K^{0}\bar{K}^{0}\to \pi^{0}\eta} &= -V_{K^{+}K^{-}\to\pi^{0}\eta}.\\
\end{aligned}
\end{equation}

Finally, the double-differential distribution of the $B^0$ decay can be written as
\begin{equation}\label{Eq:b21}
\begin{aligned}
\frac{1}{\Gamma_{B^0}}&\frac{d^2 \Gamma^\prime}{d m_{\text{inv}}(\pi a_0) d m_{\text{inv}}(\pi \eta)}= \frac{C^2}{\Gamma_{B^0}}\frac{g^2}{(2\pi)^5} \\
&\times \frac{{|\vec{p}_{D}||\vec{p}_{D}^{\,\prime}|^2|\vec{k}|^3|\vec{\tilde{q}}_{\eta}|}}{12 m_{B}^2}|{\tilde{t}_T}\times t_{K^+ K^-\to \pi \eta}|^2. \\
\end{aligned}
\end{equation}

\section{results}
\label{sec:res}

\begin{figure}[t]
\centering
\includegraphics[scale=0.6]{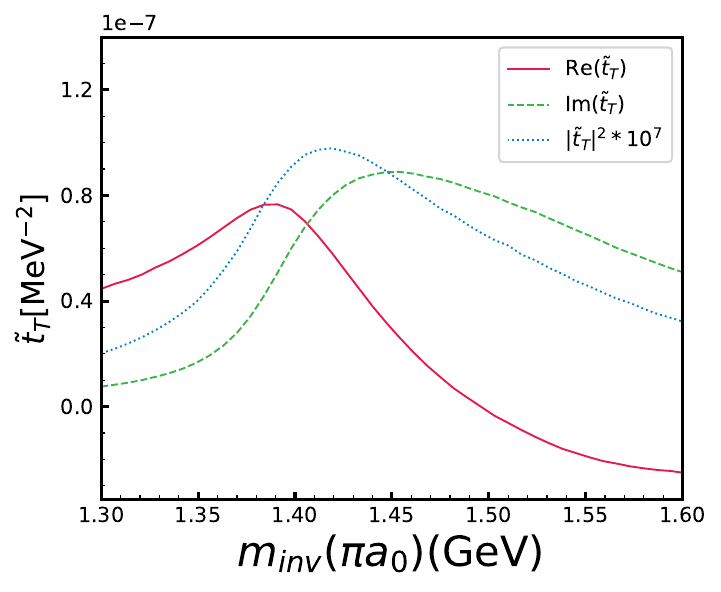}
\includegraphics[scale=0.6]{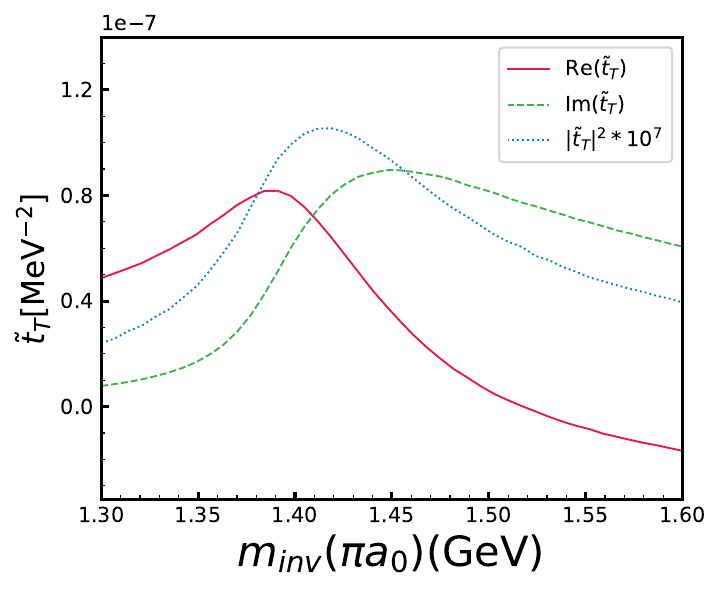}
\includegraphics[scale=0.6]{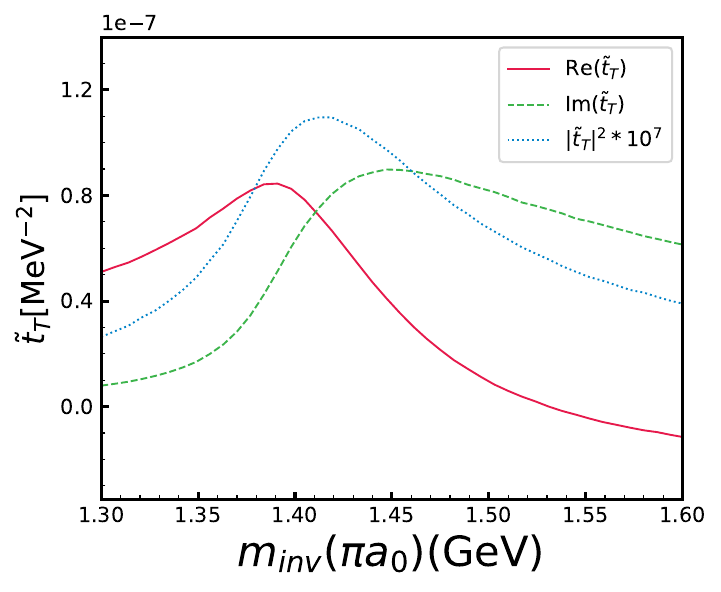}
\caption{
The $\bar{K}^{*0} K^+ K^-$ triangle amplitudes $\tilde{t}_T$ for $q_{\text{max}}=600, 800, 1000$ MeV are shown from top to bottom, respectively. 
We set $m_{a_0}=980$ MeV, and $t_T^2$ is multiplied by $10^{7}$.
}
\label{fig:tT}
\end{figure}

\begin{figure*}[t]
\centering
\includegraphics[scale=0.475]{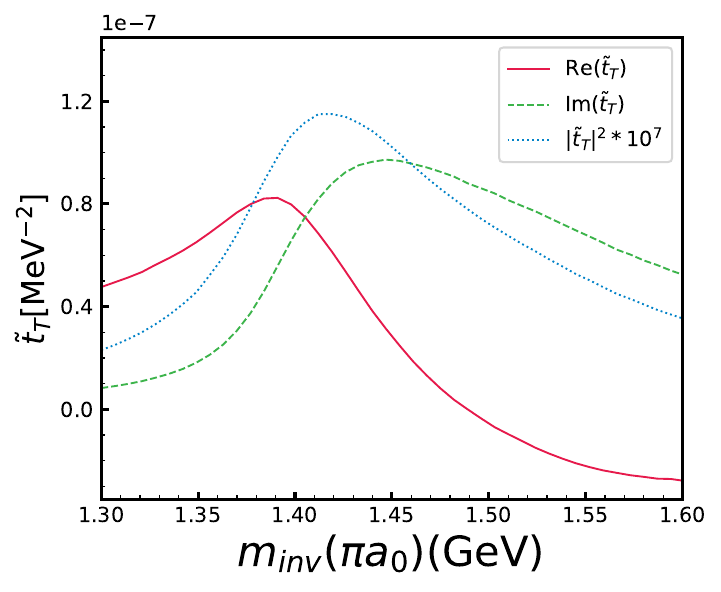}
\includegraphics[scale=0.475]{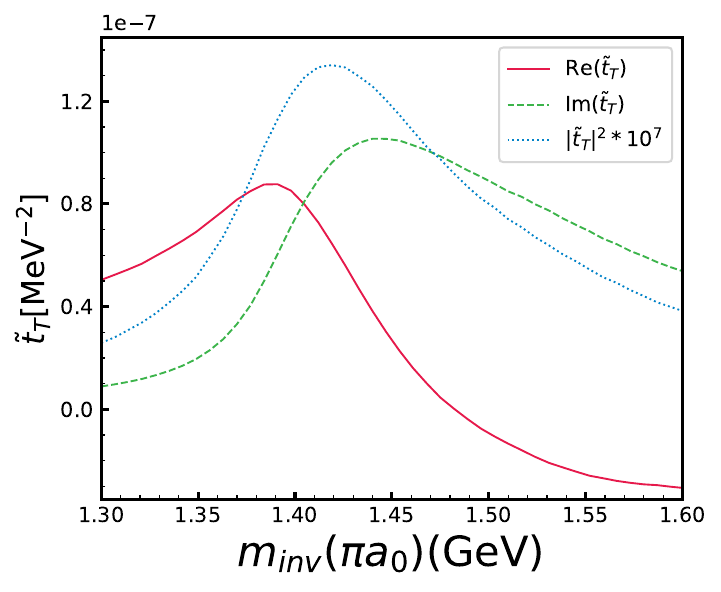}
\includegraphics[scale=0.475]{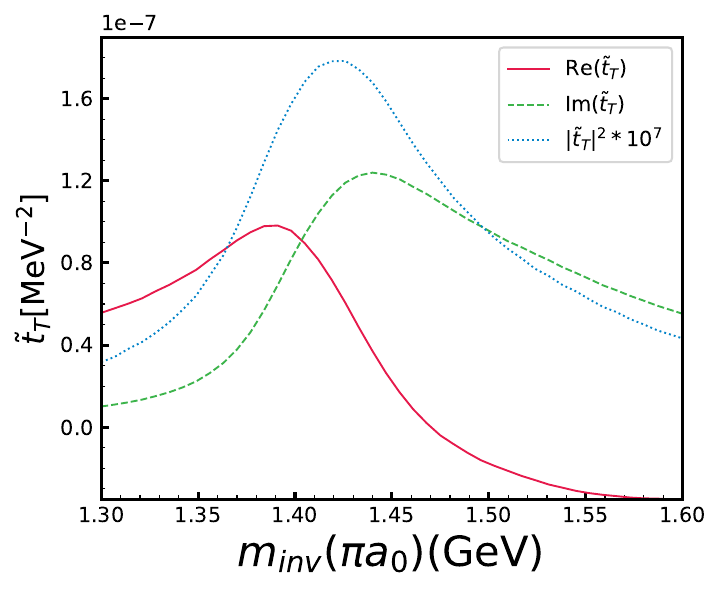}
\caption{
As in Fig.~\ref{fig:tT}, the triangle amplitudes for $m_{a_0}=983$, $985$, and $987$ MeV with $q_{\text{max}}=600$ MeV are shown.
}
\label{fig:tTa0}
\end{figure*}

In Figure~\ref{fig:tT}, we first present the triangle amplitudes $\tilde{t}_T$, $\mathrm{Im}(\tilde{t}_T)$, $\mathrm{Re}(\tilde{t}_T)$, and $|\tilde{t}_T|^2 \times 10^{7}$ as functions of the invariant mass $m_{\rm inv}(\pi a_0)$, with the $a_0(980)$ mass fixed at $m_{a_0} = 980~\mathrm{MeV}$.
The results for $|\tilde{t}_T|^2$ exhibit a clear peak around $1420~\mathrm{MeV}$, consistent with the condition in Eq.~\ref{Eq:2-1}.
The peaks of $\mathrm{Im}(\tilde{t}_T)$ and $\mathrm{Re}(\tilde{t}_T)$ appear at $1440~\mathrm{MeV}$ and $1390~\mathrm{MeV}$, respectively, arising from the TS and the $\bar{K}^{*0} K^+$ threshold.
From top to bottom, the results correspond to the cutoff parameters \replyhe{$q_{\rm max} = 600, 800$ and $1000~\mathrm{MeV}$}, respectively.
As $q_{\rm max}$ increases, the peak positions in $\mathrm{Im}(\tilde{t}_T)$, $\mathrm{Re}(\tilde{t}_T)$, and $|\tilde{t}_T|^2$ remain unchanged, while their magnitudes generally increase.
The triangle amplitudes for different masses of $a_0(980)$ are shown in Figure~\ref{fig:tTa0}.
The results indicate that the peaks remain near $1420~\mathrm{MeV}$, while the strength of the triangle amplitude increases as $m_{a_0}$ increases.
Next, we examine the behavior of the triangle amplitudes $\mathrm{Im}(\tilde{t}_T)$ and $\mathrm{Re}(\tilde{t}_T)$ as functions of the internal particle width.

\begin{figure}[t]
\centering
\includegraphics[scale=0.6]{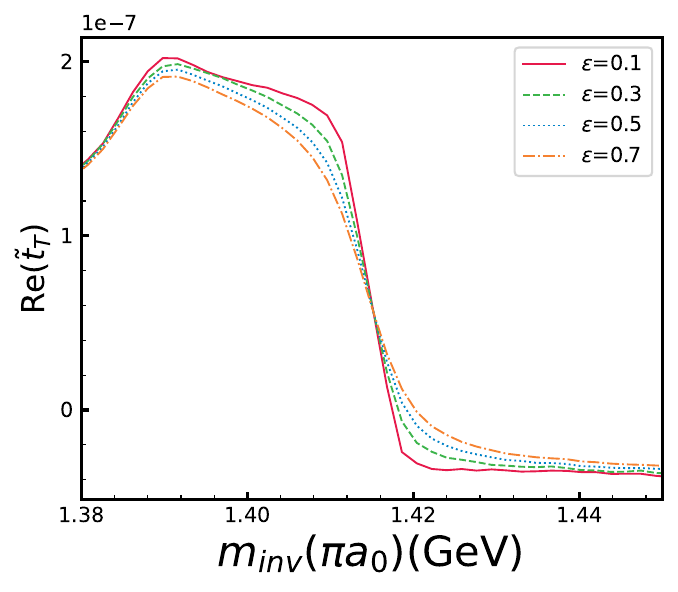}
\includegraphics[scale=0.6]{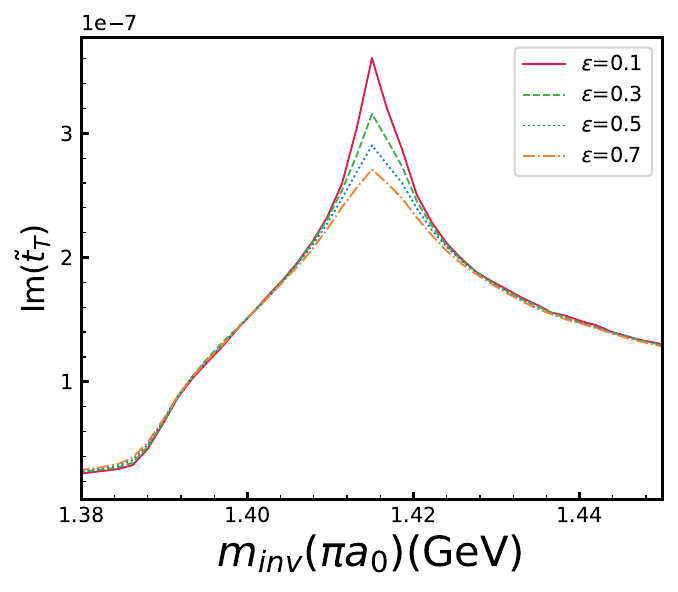}
\caption{
The dependence of the triangle amplitude Re$(\tilde{t}_T)$ and Im$(\tilde{t}_T)$ on the widths of the internal particles is shown in the upper and lower panels, respectively. 
We set $m_{a_0}=990$ MeV, which is slightly larger than the $K^- K^+$ threshold, and use $\Gamma_{\bar{K}^{*0}}/2=\epsilon=0.1,\ 0.3,\ 0.5$, and $0.7$ MeV.}
\label{fig:ReImtT}
\end{figure}

Following Ref.~\cite{Sakai:2017hpg}, FIG.~\ref{fig:ReImtT} illustrates the development of the TS by fixing $\Gamma_{K^*}/2 = \epsilon$ at different finite values close to zero.
The formation of the TS in the $\bar{K}^{*0}$-$K$-$K$ triangle loop requires the $a_0(980)$ mass to be slightly above the $K^- K^+$ threshold; therefore, we take $m_{a_0} = 990~\mathrm{MeV}$.
The distinct origins of the peaks in $\mathrm{Re}(\tilde{t}_T)$ and $\mathrm{Im}(\tilde{t}_T)$ indicate that these two components exhibit different behaviors, as shown in the upper and lower panels of FIG.~\ref{fig:ReImtT}.
Specifically, $\mathrm{Re}(\tilde{t}_T)$ exhibits a cusp at the $\bar{K}^{*0} K$ threshold around $1390~\mathrm{MeV}$, followed by a sharp decrease near the TS at $1420~\mathrm{MeV}$.
In contrast, $\mathrm{Im}(\tilde{t}_T)$ develops a narrow peak at $1420~\mathrm{MeV}$, directly reflecting the emergence of the triangle singularity.
FIG.~\ref{fig:ReImtT} also shows that, as $\epsilon$ decreases, the decrease in $\mathrm{Re}(\tilde{t}_T)$ near $1420~\mathrm{MeV}$ becomes steeper, the cusp near $1390~\mathrm{MeV}$ becomes more pronounced, and the peak in $\mathrm{Im}(\tilde{t}_T)$ becomes increasingly sharp.
In the limit $\Gamma_{K^*}/2 = \epsilon = 0$, the peak of $\mathrm{Im}(\tilde{t}_T)$ develops into a genuine singularity.

\begin{figure}[!ht]
\centering
\includegraphics[scale=0.6]{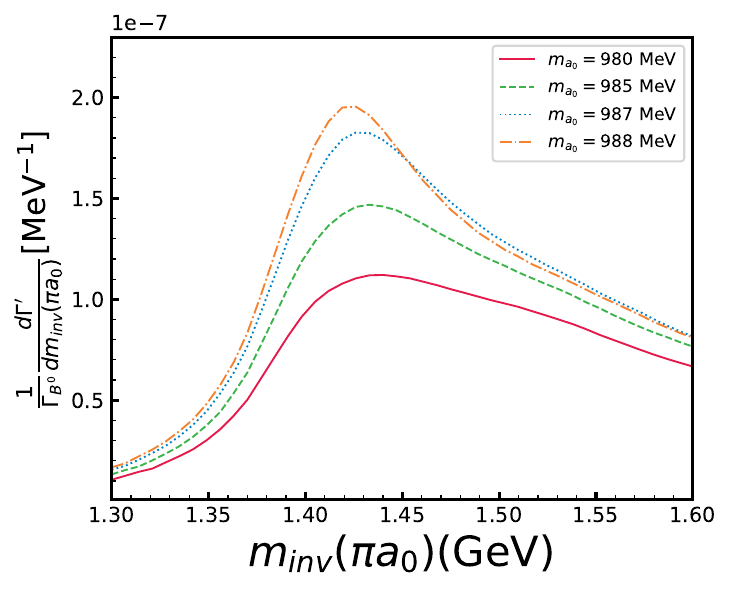}
\includegraphics[scale=0.6]{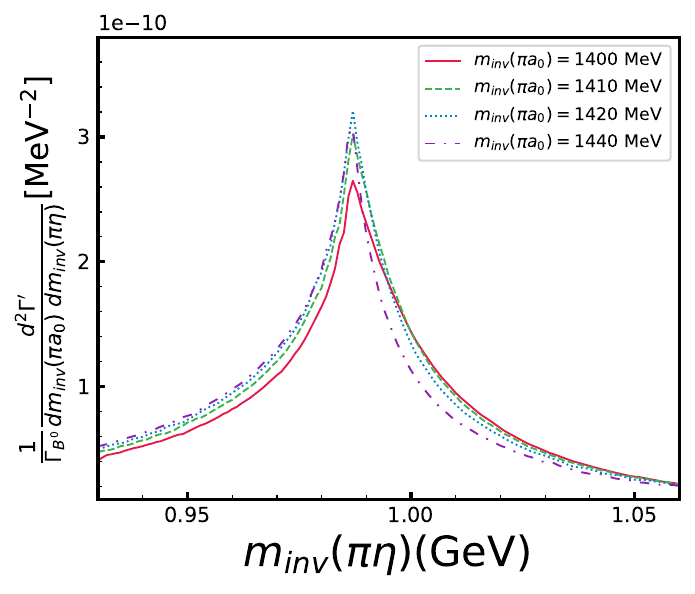}
\caption{
The upper and lower panels show, respectively, the differential distribution $\frac{d \Gamma^\prime}{d m_{\text{inv}(\pi a_0)}}$ (Eq.~\ref{Gamma-1}) as a function of the invariant mass $m_{\text{inv}}(\pi a_0)$ and the double differential distribution $\frac{1}{\Gamma_{B^0}}\frac{d^2 \Gamma^\prime}{d m_{\text{inv}}(\pi a_0) d m_{\text{inv}}(\pi \eta)}$ (Eq.~\ref{Eq:b21}) as a function of the invariant masses $m_{\text{inv}}(\pi a_0)$ and $m_{\rm inv}(\pi \eta)$.
}
\label{fig:Mpia01}
\end{figure}

\begin{figure}[t]
\centering
\includegraphics[scale=0.6]{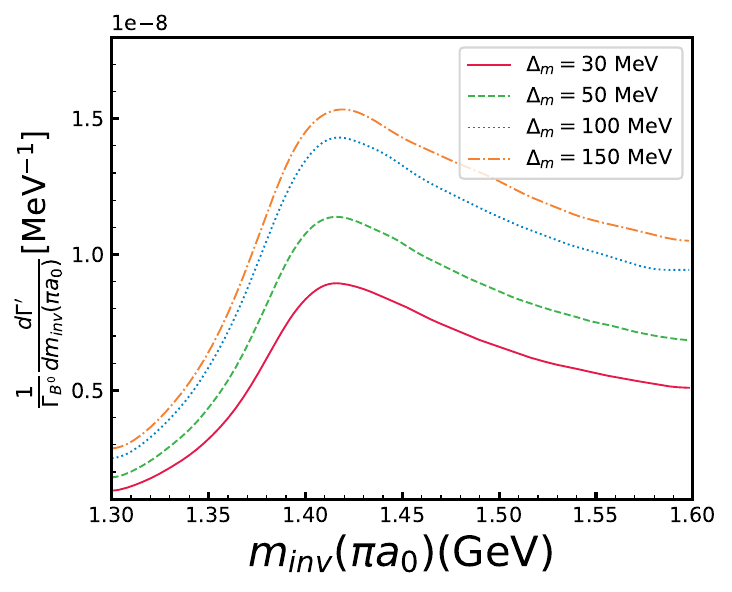}
\includegraphics[scale=0.6]{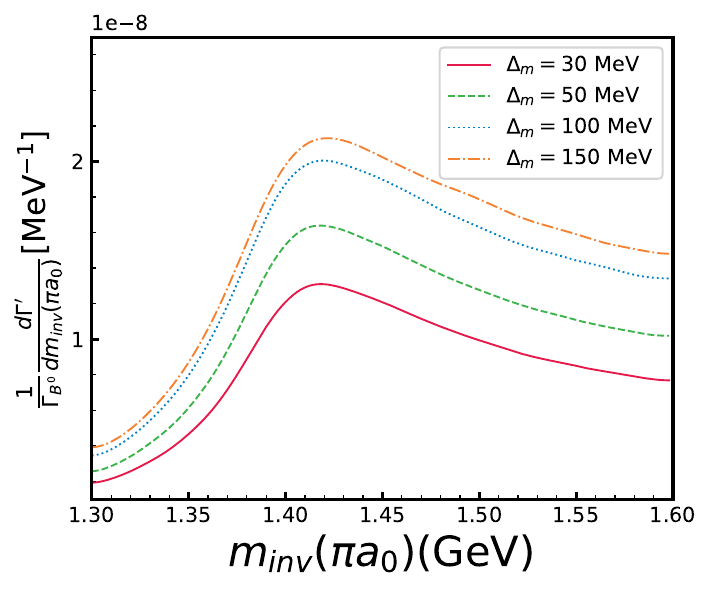}
\includegraphics[scale=0.6]{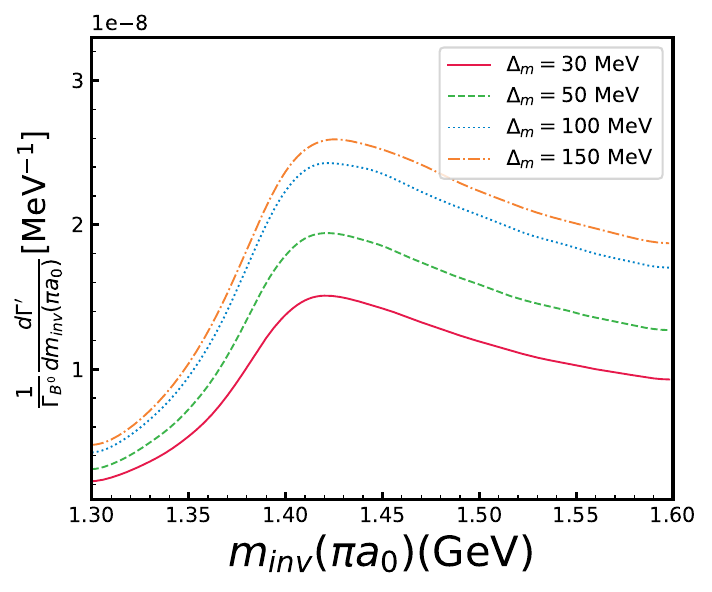}
\caption{
The differential decay width $\frac{1}{\Gamma_{B^0}}\frac{d^2 \Gamma^\prime}{d m_{\text{inv}}(\pi a_0) d m_{\text{inv}}(\pi \eta)}$ is shown as a function of the invariant mass $m_{\rm inv}(\pi a_0)$, integrated over $m_{\mathrm{inv}}(\pi\eta) \in [980~\mathrm{MeV} - \Delta_m,\; 980~\mathrm{MeV} + \Delta_m]$. From top to bottom, the curves correspond to $q_{\mathrm{max}} = 600, 800$ and $1000~\mathrm{MeV}$, respectively.
}
\label{fig:Mpia02}
\end{figure}

In Fig.~\ref{fig:Mpia01}, the differential distributions $\frac{d \Gamma^\prime}{d m_{\text{inv}(\pi a_0)}}$, Eq.~\ref{Gamma-1}, and $\frac{1}{\Gamma_{B^0}}\frac{d^2 \Gamma^\prime}{d m_{\text{inv}}(\pi a_0) d m_{\text{inv}}(\pi \eta)}$, Eq.~\ref{Eq:b21}, are shown as functions of the invariant masses $m_{\text{inv}}(\pi a_0)$ and $m_{\rm inv}(\pi \eta)$ in the upper and lower panels, respectively.
In the upper panel of Fig.~\ref{fig:Mpia01}, the differential mass distribution for the $B^0 \to D^- \pi^+ a_0(980)$ decay with respect to $m_{\rm inv}(\pi a_0)$ is shown for fixed values \replyhe{ $m_{\rm inv}(\pi \eta) = 980, 985, 987 $ and $ 988~\mathrm{MeV}$} with $q_{\rm max} = 600~\mathrm{MeV}$.
It is evident that increasing $m_{a_0}$ enhances the decay width of $B^0$, while the peak position remains at the TS, $m_{\rm inv}(\pi a_0) \simeq 1420~\mathrm{MeV}$.
In the lower panel of Fig.~\ref{fig:Mpia01}, the dependence of $\frac{1}{\Gamma_{B^0}}\frac{d^2 \Gamma^\prime}{d m_{\text{inv}}(\pi a_0) d m_{\text{inv}}(\pi \eta)}$ on $m_{\rm inv}(\pi \eta)$ is shown for fixed values \replyhe{ $m_{\rm inv}(\pi a_0) = 1400, 1410, 1420$ and $1440~\mathrm{MeV}$} around the TS.
A pronounced peak appears around $980~\mathrm{MeV}$, indicating that the main contribution to the decay width comes from the region $m_{\rm inv}(\pi \eta) \simeq 980~\mathrm{MeV}$.
Moreover, the peak for $m_{\rm inv}(\pi a_0) = 1420~\mathrm{MeV}$, close to the TS, is significantly sharper than those at the other values.
Therefore, when focusing on the $a_0(980)$ region, we can integrate the double differential distribution $\frac{1}{\Gamma_B} \frac{d^2 \Gamma^\prime}{d m_{\rm inv}(\pi a_0)\, d m_{\rm inv}(\pi \eta)}$ over $m_{\rm inv}(\pi \eta)$ around $m_{\rm inv}(\pi \eta) = 980~\mathrm{MeV}$.
In Fig.~\ref{fig:Mpia02}, we present the results obtained by integrating over $m_{\mathrm{inv}}(\pi\eta) \in [980~\mathrm{MeV} - \Delta_m,\; 980~\mathrm{MeV} + \Delta_m]$ with $\Delta_m = 30,\,50,\,100$ and $150~\mathrm{MeV}$.
The results show a clear peak at $1420~\mathrm{MeV}$ when the $a_0(980)$ is considered a dynamically generated state, consistent with the results in Fig.~\ref{fig:Mpia01}.
A comparison of the subfigures for $q_{\rm max} = 600, 800$ and $1000~\mathrm{MeV}$ shows that the decay width exhibits a slight enhancement as the cutoff parameter increases.

\section{summary}
\label{sec:sum}
In this work, we investigate the triangle mechanism in the $B^0 \to D^- \pi^+ a_0(980){(\pi^0 \eta)}$ decay and explore the nature of $a_0(980)$ as a dynamically generated state arising from meson-meson interactions.
The $\bar{K}^{*0}$-$K^+$-$K^-$ triangle loop originates from the weak decay $B^0 \to D^- K^+ \bar{K}^{*0}$, followed by $\bar{K}^{*0} \to \pi^+ K^-$ and the subsequent rescattering of $K^+ K^-$ to form $a_0(980)$.
We provide analytical expressions for the differential decay widths $\frac{d\Gamma}{d m_{\rm inv}(\pi a_0)}$ and $\frac{d^2 \Gamma}{d m_{\rm inv}(\pi a_0)\, d m_{\rm inv}(\pi \eta)}$ corresponding to the three-body and four-body final states, respectively, including the $\bar{K}^{*0}$-$K^+$-$K^-$ triangle loop amplitude.
The triangle amplitude produces a pronounced peak near $1420~\mathrm{MeV}$ in the $\pi a_0(980)$ invariant mass distribution, which can be interpreted as a TS signal.
The dependence of the TS on the internal $\bar{K}^{*0}$ width is also investigated.
The \replyhe{$K^+K^-$ rescattering} generates a pronounced peak near $980~\mathrm{MeV}$ in the $\pi^0\eta$ invariant mass distribution, reflecting the formation of the $a_0(980)$.
Meanwhile, the invariant mass spectrum of the $\pi^+\pi^0\eta$ system exhibits a peak around $1420~\mathrm{MeV}$, consistent with the results obtained from the three-body final state.
Based on the current observations and analyses by the \replyhe{COMPASS Collaboration}, we expect that \replyhe{$B^0$ decays can serve} as a potential platform for further confirming the \replyhe{TS nature of the $a_1(1420)$} in future experiments.

\acknowledgments
D.Z.H. is supported by the Heze University doctoral fund project No.~010008002039037.
X.L. is supported by the National Natural Science Foundation of China under Grant No.~12205002.

\bibliography{refv2}
\end{document}